\documentclass[prb, twocolumn]{revtex4}
\usepackage{graphicx}
\usepackage{latexsym}
\usepackage{amsmath}
\usepackage{amssymb}
\usepackage{epstopdf}
\usepackage{enumerate}
\usepackage{setspace}  
\begin{document}
\title{Quantum critical point shifts under superconductivity: \\
the pnictides  and the cuprates }

\author{Eun Gook Moon and Subir Sachdev}
\affiliation{Department of Physics, Harvard University, Cambridge MA
02138}
\date{\today}

\begin{abstract}
We compare the position of an ordering transition in a metal to that in a 
superconductor. For the spin density wave (SDW) transition, we find that the quantum critical point
shifts by order $|\Delta|$, where $\Delta$ is pairing amplitude, so that the region of SDW order
is smaller in the superconductor than in the metal. This shift is larger than the $\sim |\Delta|^2$ shift
predicted by theories of competing orders which ignore Fermi surface effects. For Ising-nematic order,
the shift from Fermi surface effects
remains of order $|\Delta|^2$. We discuss implications of these results for the phase diagrams
of the cuprates and the pnictides. We conclude that recent observations imply that the Ising-nematic order is
tied to the square of the SDW order in the pnictides, but not in the cuprates.
\end{abstract}

\maketitle

\section{Introduction}
\label{intro}

The interplay between spin density wave (SDW) ordering and superconductivity clearly plays a central role
in the physics of a variety of quasi two-dimensional correlated electron materials. 
This is evident from recent studies of the phase diagram of the ferro-pnictides \cite{imai1,joerg0,imai2,imai3,joerg1,joerg2,nmrjapan} and
the `115' family of heavy-fermion compounds \cite{knebel}. In the cuprates, it has been argued that
$d$-wave superconductivity is
induced by SDW fluctuations in a metal \cite{scalapino}, and this has been the starting point for numerous
studies of the complex phase diagram \cite{ChubukovLong,max2}. In all these materials, there is a regime
of co-existence between SDW ordering and superconductivity, and this opens the way to a study
of the `competition' between these orders \cite{sachdevzhang}: this competition can be tuned by an applied magnetic
field, as has been studied in a number of revealing experiments \cite{katano,lake,jtran2,boris,chang1,chang2,mesot3,keimer} 
on the LSCO and YBCO series of superconductors.

\begin{figure}[t]
\includegraphics[width=3.0 in]{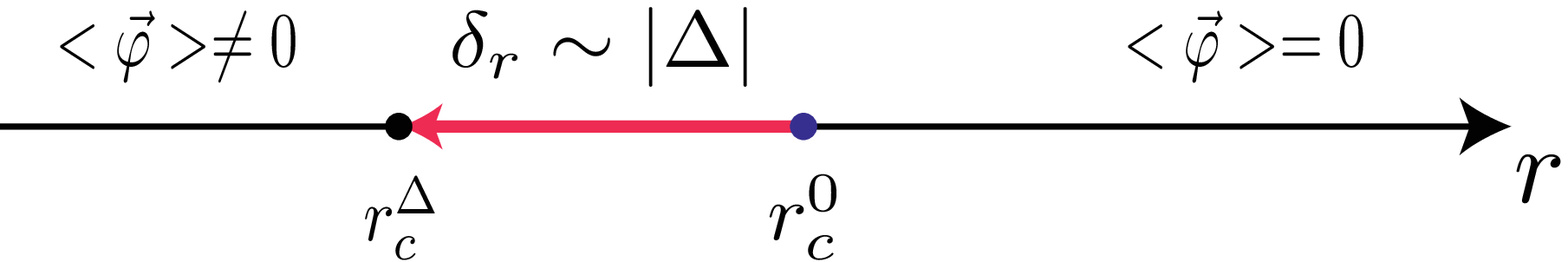}
\includegraphics[width=3.0 in]{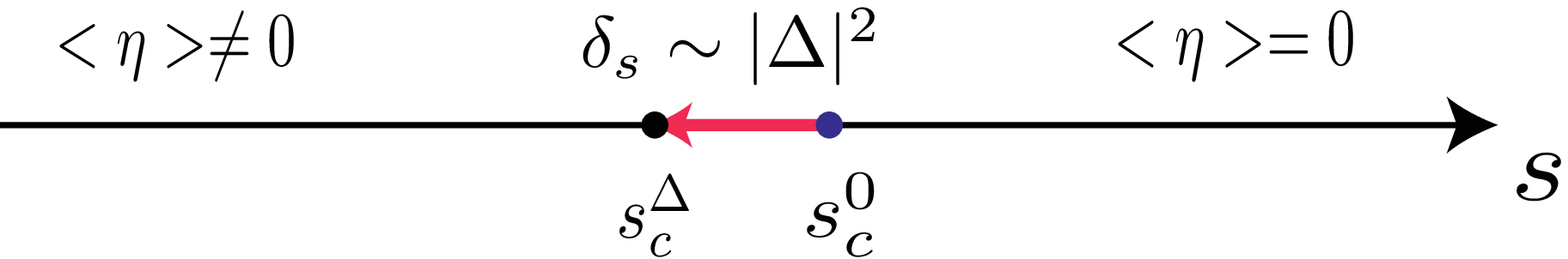}
\caption{ The SDW and nematic critical points shifts from Fermi surface effects. 
The tuning parameters $r,s$ are for the SDW and the nematic phase transitions. The critical points in the metal are at $r_c^{0}, s_c^{0}$, and under superconductivity, these
shift to $r_c^{\Delta}, s_c^{\Delta}$, towards the ordered phases. }
\label{totalshift}
\end{figure}

This paper will discuss a question that arises naturally in the study of such competing orders \cite{eg,qcnp}.
We consider, first, the `parent' quantum critical point as that associated with the onset of SDW order, $\vec{\varphi}$,
in a metal. To access this point we have to suppress superconductivity in some manner, say by the application 
of a magnetic field. This parent critical point will occur at a value $r_c^0$ of some tuning parameter $r$, which could
be the carrier concentration or the applied pressure. We define $r$ so that $r < r_c^0$ is the SDW phase with $\langle
\vec{\varphi} \rangle \neq 0$; see Fig.~\ref{totalshift}.
The value of $r_c^0$ is clearly material specific, and will
depend upon numerous microscopic details. Then, we turn our attention to the onset of SDW order within
the superconductor (SC); we characterize the latter by a gap amplitude $\Delta$, and denote the critical
value of $r$ by $r_c^\Delta$. The essence of the picture of competing orders is that the onset of superconductivity
should shrink the region of SDW order, and hence $r_c^0 > r_c^\Delta$. We will be interested here in particular
in the magnitude of the shift $r_c^0 - r_c^\Delta$. We will see that the shift is dominated by low energy physics, and so has
a universal character.
This shift $r_c^0 - r_c^\Delta$ played a central role in the phase diagrams presented in Refs.~\onlinecite{qcnp,zhang}, and
applied to the cuprates. Recent work has shown that similar phase diagrams also apply to the pnictides \cite{joerg1,joerg2} and the 115 compounds \cite{knebel}. In the pnictides, a ``backbending'' of the onset of SDW order upon entering
the SC phase, consistent with the idea of $r_c^0 - r_c^\Delta > 0$.

Let us begin by computing the shift $r_c^0 - r_c^\Delta$ in mean-field Landau theory. The simplest free energy of the SDW
and SC order parameters has the form \cite{zhang}:
\begin{equation}
\mathcal{L} = (r - r_c^0) \vec{\varphi}^2 + u \left (\vec{\varphi}^2 \right)^2 + \tilde{r} |\Delta|^2
+ \tilde{u} |\Delta|^4 + \kappa \vec{\varphi}^2 |\Delta|^2 . \label{landau}
\end{equation}
Here $\kappa > 0$ is the phenomenological parameter which controls the competition between the order parameters.
Examining the onset of a phase with $\langle \vec{\varphi} \rangle \neq 0$ in the superconductor with $\Delta \neq 0$,
we conclude immediately from Eq.~(\ref{landau}) that
\begin{equation}
  r_c^0 - r_c^\Delta = \kappa |\Delta |^2 .
\label{landaushift}
\end{equation}
Such a shift was a key feature of the theory in Ref.~\onlinecite{zhang}.

The primary focus of the previous work was in the lower field region, where the superconductivity is well-formed,
and $\Delta$ is large.
Here it is appropriate to treat the superconductivity in a mean-field manner, and ignore pairing fluctuations,
while treating spin fluctuations more carefully. 
The present paper turns the focus to higher fields, where eventually superconductivity is lost. Here, clearly, 
Landau theory cannot be expected to apply to the superconducting order. Moreover, we expect the Fermi surface
of the electrons to be revealed, and a more careful treatment of the electronic degrees of freedom is called for. 
One of the primary results of our paper will be that the Landau theory result in Eq.~(\ref{landaushift}) breaks down
for small $\Delta$, and in particular in the limit
$|\Delta| \rightarrow 0$.
This is a consequence of the crucial importance of Fermi surface physics in determining the position of the SDW
transition at $T=0$. Instead, we will show from the physics of the ``hot spots'' on the Fermi surface that
the shift is larger, with
\begin{equation}
r_c^0 - r_c^\Delta \sim C |\Delta|.
\label{sdwshift}
\end{equation}
For the competing order picture to hold, we require that $C>0$. Somewhat surprisingly, we will find that
our results for $C$ are not transparently positive definite. Different regions of the Fermi surface contribute
opposing signs, so that determining the final sign of $C$ becomes a delicate computation.
In particular $C$ will depend upon the vicinity of `hot spots' on the Fermi surface, which are special points
connected by the SDW ordering wavevector. We will find that the immediate vicinity of the hot
spots contributes a positive sign to $C$, while farther regions contribute a negative sign. Thus the primary
competition between SDW and superconductivity happens at the hot spots, while other regions
of the Fermi surface which survive the onset of SDW order continue to yield an attraction between
SDW and superconductivity. For the case where the two hotspots connected by the SDW ordering wavevector
are equivalent under a lattice symmetry operation ({\em i.e.\/} they have the same pairing gap
and the same magnitude of the Fermi velocity), we will find that distinct contributions to $C$ exactly
compensate each other, so that $C=0$. However, in the case that the two spots are not crystallographically equivalent
(which is the generic situation in both the cuprates and the pnictides), we will show that $C>0$.
A positive $C$ is indicated in Fig.~\ref{totalshift}.

We had considered the shift in SDW ordering due to superconductivity in a previous work \cite{eg}.
However, in that work, the metallic and superconducting states were not Fermi liquids and BCS states respectively,
but rather fractionalized states known as `algebraic charge liquids' \cite{rkk1,rkk2,victor,subir}.
In this case, we found that the competition between SDW and superconductivity was robust, and
always yielded a shrinking in the size of the SDW region.
We will not consider such exotic states here, but work entirely within the framework of Fermi liquid theory,
in which the onset of superconductivity leads to a traditional BCS superconductor. In this context the
interplay between SDW and SC in a Fermi liquid is conveniently encapsulated in the 
`spin-fermion' model \cite{abanov}. We will find the same qualitative shift in the SDW critical point
as found earlier \cite{eg}, and the  estimate in Eq.~(\ref{sdwshift}).

We will also generalize our methods to analyze the shifts in the quantum critical points of
other orderings between the metallic and superconducting phases. Specifically, we will consider
charge density wave (CDW) order and Ising-nematic order, $\eta$. 
We will find that the CDW shift initially appears to be formally
similar to Eq.~(\ref{sdwshift}), but the coefficient $C$ is found to be exactly zero; terms higher order
in $\Delta$ do indicate competition with superconductivity, but 
the CDW critical point shift is much smaller than the SDW's 
For Ising-nematic
order, $\eta$, we will also find that the Fermi surface 
result is similar to that in Landau theory, as in Eq.~(\ref{landaushift}).
This smaller shift is also illustrated in Fig.~\ref{totalshift}. The weaker effect upon $\eta$ is due to
its reduced sensitivity to the gap opened by superconductivity on the Fermi surface. 

We will present a detailed discussion of the implication of these results for the pnictide and cuprate phase diagrams to 
Section~\ref{conclusion}. However, let us highlight here an important inference that will follow
from our computations. We will argue that the experimental phase diagrams imply
that Ising-nematic ordering and SDW ordering are independent instabilities of the Fermi surface 
for the cuprates. In contrast, for the pnictides, our conclusion will be that the SDW ordering
is the primary Fermi surface instability, and the Ising-nematic ordering is a secondary response
to the square of the SDW order parameter.

This paper is structured as follows. 
We will begin in Section~\ref{spin-fermion} by introducing our starting point, the spin-fermion model with the SDW fluctuation. 
Within the spin-fermion theory, we show that the SDW fluctuation induces the $d$ wave pairing for the cuprate and $s_{+-}$ for the pnictides instead usual $s$ wave pairings in Section \ref{instabilities}. 
Assuming the $d$ or $s_{+-}$ wave pairings in each case, we extend the spin-fermion theory into the theory with pairing and other possible orders in Section~\ref{theory}.
In section \ref{QCPshift}, we show the quantum critical point shifts toward the ordered phase, which explicitly shows the competition between superconductivity and the SDW phase. Section~\ref{conclusion} presents our conclusions.

\section{The spin-fermion model} \label{spin-fermion}
\begin{figure}
\includegraphics[width=2.0 in]{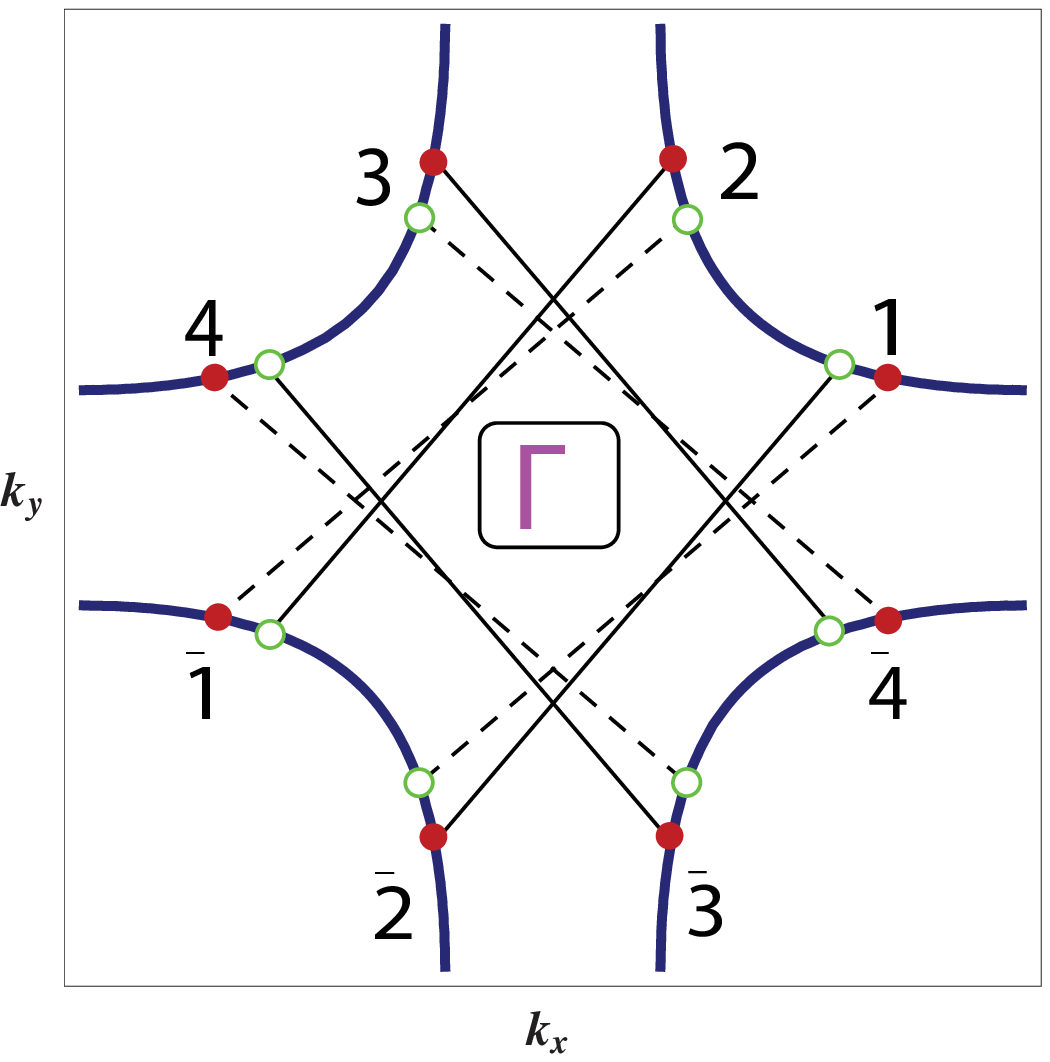}
\includegraphics[width=2.0 in]{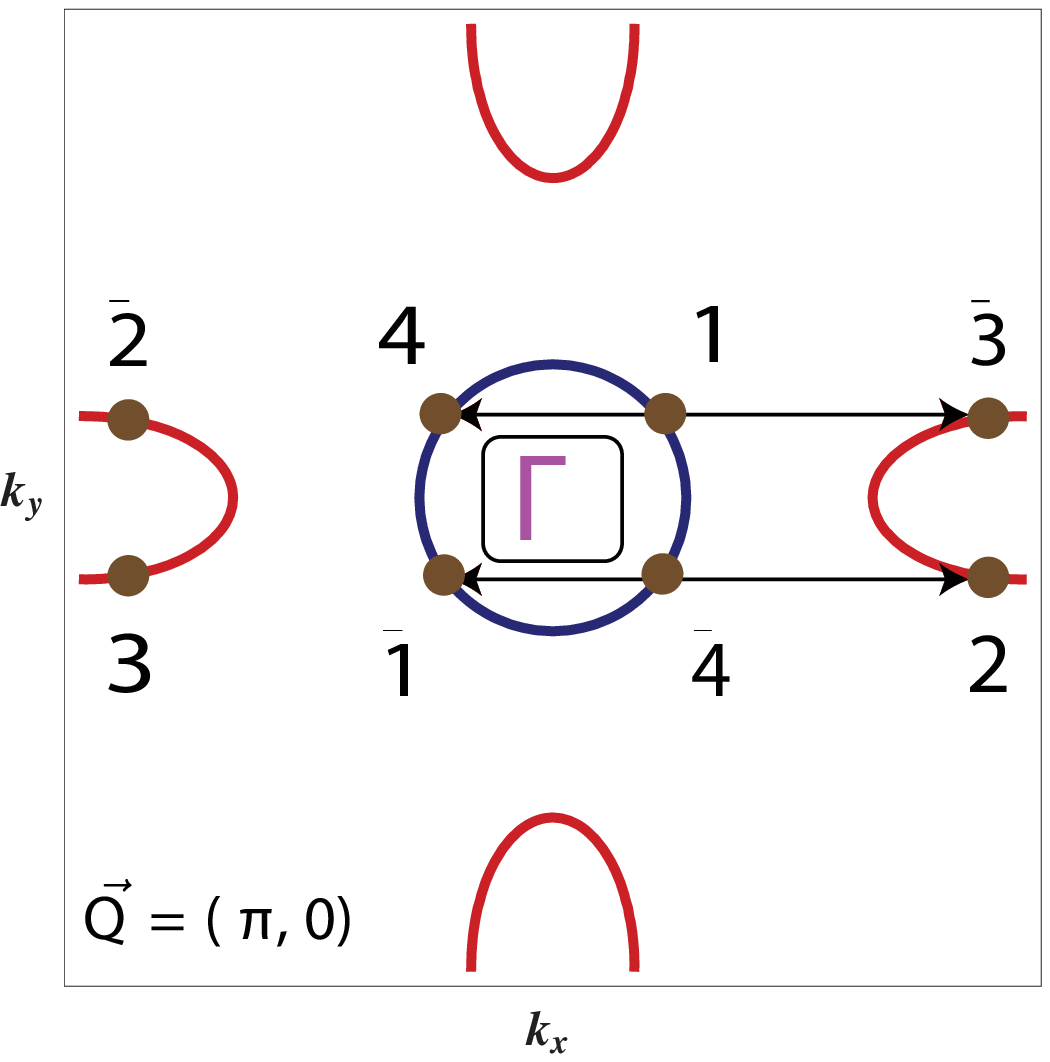}
\caption{ Hot spots near SDW transition. In the upper and lower panel, large Fermi surface for the cuprates and two band structure for the pnictides are shown. The ordering vectors for the cuprates are $\vec{Q}_1 = \frac{2 \pi}{a_0} (\frac{1}{2} - \vartheta, \frac{1}{2}), ~ \vec{Q}_2 =\frac{2 \pi}{a_0} (\frac{1}{2}, \frac{1}{2} - \vartheta))$, and the pnictides have $\vec{Q} = (\pi,0)$.
In the upper panel, the filled and empty circles have different distances from the node, which means the gap magnitudes are different. Two ordering wave vectors are represented by the dashed and thin lines in the upper panel and arrowed lines are in the bottom. For the pnictides, the electron band is distorted because there is no symmetry that guarantees identicalness of the hole and the electron band, which can also induces different gap functions.\cite{gapan1,gapan2} In both panels, every hot spot is numbered and the bar notation is used for the negative. }
\label{hotspots}
\end{figure}

We will study the system with SDW quantum phase transition in two dimensional system. 
The main ingredients of the spin-fermion model are  Fermi surfaces and the SDW order parameter.
Let us first consider generic microscopic Hamiltonians for the cuprates and the pnictides. 
\begin{eqnarray}
H_{Cu} &=& \sum_{k} \epsilon_c (k) c^{\dagger}_a c_a +H_{sdw}\nonumber \\
H_{Fe} &=& \sum_{k} \epsilon_c (k) c^{\dagger}_a c_a +  \sum_{k} \epsilon_d (k) d^{\dagger}_a d_a +H_{sdw}
\end{eqnarray}
Here, we consider ``minimal Fermi surfaces'' for both materials, where $c$ correspond to the one band electron for the cuprates around the $\Gamma$ point, and for the pnictides,  $c,d$ correspond to the hole and electron bands centered at $(0,0)$ and $(\pm \pi/ a_0,0)$. 
All terms containing the SDW operators are in $H_{sdw}$.
In Fig. \ref{hotspots}, we illustrate typical hole-doping cuprates  large Fermi surface and pnictides' two band Fermi surfaces.
To see general features, we consider incommensurate ordering wave-vectors for cuprates\cite{keimer},
\begin{eqnarray}
\vec{Q}_1 = \frac{2 \pi}{a_0} \left(\frac{1}{2} - \vartheta, \frac{1}{2} \right), ~ 
\vec{Q}_2 = \frac{2 \pi}{a_0} \left(\frac{1}{2}, \frac{1}{2} - \vartheta \right).
\end{eqnarray}
For the pnictides, we represent the ordering wavevector as $\vec{Q} =( \pi /a_0, 0)$ explicitly in the figure, but there is another hot spot with the ordering vector, $\vec{Q} = (0,\pi / a_0)$. 
From now on, we set the lattice constant as a fundamental unit as usual. 
As it is well-known in the literature,\cite{zhang} the incommensurate SDW fluctuation is decribed by two complex vector wave functions. 
\begin{eqnarray}
\vec{S} = Re[ \vec{\Phi}_1e^{i \vec{Q}_1 \cdot \vec{r}}  +  \vec{\Phi}_2 e^{i \vec{Q}_2 \cdot \vec{r}} ]
\end{eqnarray}
In the figure, two distinct hot spots are represented by filled and empty circles in the cuprate Fermi surface. 
The filled one is farther from the nodal point than the empty one. Note that the incommensurate SDW fluctuation links a filled circle with a empty circle. 
If we consider one special case, the commensurate SDW fluctuation, two kinds of hot spots become identical, and we only need one real $O(3)$ field to describe the SDW fluctuation as usual.
  
Because the hot spots mainly contribute to the transition in the low energy theory, we write the electron annihilation operator as combination of the hot spot continuum fields as follows. 
For the cuprate, 
\begin{eqnarray}
c_{a}(x) &\sim& \sum_{j} f_{j, a} (x) e^{i k_{f,j} \cdot x} + g_{j, a} (x) e^{i k_{g,j} \cdot x} \nonumber \\
&& \quad, \quad j=(1, \cdots, 4, \bar1 , \cdots, \bar 4). 
\end{eqnarray}
The $f, g$ fields represent the nearer and farther fields from the nodal points, so the SDW fluctuation mixes the $f$ and $g$ fields. 
The commensurate limit means $f$ and $g$ fields become identical. 
For the pnictides case, two bands can be described by hot spot fields as follows. 
\begin{eqnarray}
c_a &\sim& f_{1a} e^{i K_1 \cdot x} + f_{\bar{1}a} e^{-i K_1 \cdot x} \nonumber \\
&+& f_{4a} e^{i K_4 \cdot x} + f_{\bar{4}a} e^{-i K_4 \cdot x}  \nonumber \\
d_a &\sim& g_{2a} e^{i K_2 \cdot x} + g_{\bar{2}a} e^{-i K_2 \cdot x} \nonumber \\
&+& g_{3a} e^{i K_3 \cdot x} + g_{\bar{3}a} e^{-i K_3 \cdot x}  
\end{eqnarray}
As in the cuprates, the SDW fluctuation mixes $f$ and $g$ particles in this notation. 
We note that the electron band does not have the shape of circle, which makes the SDW possible, and there is no symmetry that guarantees the identity of the two bands. 
For convenience, we  focus on the commensurate case in the remaining of this section, and the next two sections for describing spin-fermion models.
But, later in the Sec. \ref{QCPshift}, we will come back to the general incommensurate cases and consider the critical point shifts in general. 

The spin-fermion model with commensurate SDW simply becomes 
\begin{eqnarray}
{\mathcal L} &=&   \frac{1}{2}  (\partial_{\tau} \vec \varphi)^2 + \frac{1}{2}(\nabla \vec  \varphi)^2 + \frac{r}{2} ( \vec \varphi)^2 + \frac{u}{4} ( \vec \varphi^2)^2 \nonumber  \\ 
&+& f^{\dagger}_{j,a} (\partial_{\tau} - i \vec{v}_{j} \cdot \nabla - a v_f^2 \nabla^2) f_{j,a}  \nonumber \\ 
&+& \lambda \vec \varphi \left(f^{\dagger}_{j,a}  \vec{\sigma}_{a, b} f_{j',b}  \right) {\rm M}_{j,j'},
\label{lagrangian}
\end{eqnarray}
where ${\rm M}_{j,j'}$ is non-zero constant when $k_j,k_j'$ are connected by $\vec Q $. 
The first and second lines describe the dynamics of the spin and fermion sectors. 
The third line is the ``Yukawa'' coupling term. 
Note that we explicitly include the second derivative kinetic term for the fermions in the theory, which becomes irrelevant if we only focus on the SDW phase transition. 
Such term is not necessary for describing the SDW phase transition with non-collinear Fermi velocities only, 
but the existence plays an important role in extending the theory to the one with pairing and nematic orders.
Note that the final form of the spin-fermion model is exactly the same in both the cuprates and the pnictides even though the microscopic band structure and the ordering vectors are completely different. This means the physics for the SDW transition is universal and we can focus on one case and apply the result to the other case.

As usual, the effective action for the SDW phase transition is given by integrating out the fermions and expand with order parameters.
\begin{eqnarray}
{\mathcal S}_{H} &=& \int \frac{d^2 k} {4 \pi^2} T \sum_{\omega_n} \frac{1}{2}[k^2 + \gamma |\omega_n| +r -\chi_0] |\varphi_a(k,\omega_n)|^2 \nonumber \\
&& + \frac{u}{4!} \int d^2 x d \tau (\varphi_a^2(x,\tau))^2
\label{Hertz}
\end{eqnarray}
This Hertz-type theory is well-known and it describes the SDW fluctuation with dynamical critical exponent $z=2$ at least in the zeroth order. 
In this paper, we only focus on the critical point shifts rather than critical properties of the transition itself. \cite{max2}

\section{Pairing instabilities} \label{instabilities}
\begin{figure}
\includegraphics[width=1.5 in]{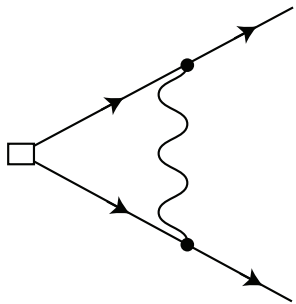}
\caption{Pairing vertex. The line with arrow is fermion and the wavy line is for the SDW fluctuation. } \label{vertex}
\end{figure}

Within the spin-fermion model, we can address pairing problems naturally.
If we consider the SDW fluctuation as a pairing boson, then we need to investigate plausibility of the pairing instabilities by the SDW. 
The basic idea is following. If we assume there is  infinitesimal pairing, then the pairing becomes enhanced or suppressed by the integrating out higher energy-momentum contributions depending on the possibility of the pairing channel. 
In the Fig. \ref{vertex} such a pairing vertex is illustrated. 
Note that the fermions with opposite momentums are paired, so the participating fermions in the pairing is not the same as ones in the SDW in general. 
If we consider the $s$ wave channel in the cuprates and the $s_{++}$ channel for the pnictides, the pairing and its vertex correction are 
\begin{eqnarray}
\Phi_s &\equiv&\varepsilon_{ab}( f_{1,a} f_{\bar1,b} + f_{2,a} f_{\bar2,b} + f_{3,a} f_{\bar 3,b} +f_{4,a} f_{\bar4,b})  \nonumber \\
g^s_{\Phi} &=& g^s_{\Phi,0} \left(1-  \lambda_{\varphi}^2 \int_{k,\omega} \frac{1}{\omega^2 + \varepsilon_k^2} \frac{1}{ \gamma |\omega| + k^2 +\xi^{-2}} \right) .~~~~
\end{eqnarray}
Note that the relative sign between hot spots does not change, which is the main characteristic of the $s$ wave pairing. 
As we can see, the vertex becomes irrelevant in the low energy limit  in the RG sense.
Therefore, the SDW fluctuation cannot mediate usual $s$ wave pairing for the cuprates and $s_{++}$ for the pnictides.

However, in the $d$ wave channel of the cuprates \cite{scalapino} and the $s_{+-}$ channel \cite{mazinpm,jiangping} of the pnictides, the relative sign between the hotspots 
changes, given the pairing symmetries. Such relative sign changes allow the pairing channel's enhancement. 
The pairing and its vertex correction are
\begin{eqnarray}
\Phi_d &\equiv&\varepsilon_{ab}( f_{1,a} f_{\bar1,b} - f_{2,a} f_{\bar2,b} - f_{3,a} f_{\bar 3,b} +f_{4,a} f_{\bar4,b}) \nonumber \\
g^d_{\Phi}&=& g^d_{\Phi,0} \left(1+ \lambda_{\varphi}^2 \int_{k,\omega} \frac{1}{\omega^2 + \varepsilon_k^2} \frac{1}{ \gamma |\omega| + k^2 +\xi^{-2}} \right) .~~~~~
\label{pairing}
\end{eqnarray}
Clearly, the alternative sign change induces enhancement of the $d$, $s_{+-}$ wave pairings in the low energy limit.
Therefore, the $d$ and $s_{+-}$ pairing is natural under the SDW fluctuations rather than usual $s$ and $s_{++}$ pairings in conventional theory.
In the next sections, we assume the existence of the pairings in each case and incorporate them into the spin-fermion model in a manner consistent with symmetry. 
Moreover, by symmetry consideration, we also introduce other possible order parameters such as a nematic order and charge density wave and extend our theory to incorporate them. 

Notice that the vertex correction in  the Eq. (\ref{pairing}) is logarithmically divergent  if we have finite correlation length.
Such behavior is  a well-known signature of the conventional BCS theory. 
However, in the quantum critical region, the pairing boson is softened and the quantum fluctuation becomes important. 
The nature of this quantum critical pairing has been discussed in Refs.~\onlinecite{abanov,max2}.

\section{Microscopic symmetry and effective theory} \label{theory}

To extend the spin-fermion model to the one with pairing terms and nematic order parameter, let us consider microscopic symmetries and thier transformations because
the square lattice symmetry should be respected in the low energy theory. 
Hereafter, we analyze the symmetry in terms of the cuprate problem unless otherwise stated. 
It is easy to extend it to the pnictides case. 
Due to the $d_{x^2-y^2}$ wave property, the pairing term's rotation and reflection needs additional factors.
In Appendix~\ref{app:symmetry}, 
we show the explicit transformation properties of fields and bilinear terms to avoid notation ambiguity.
Nambu spinors for particles are defined in a usual way:
\begin{eqnarray}
\Psi_{i,a} \equiv \begin{pmatrix} f_{i,a} \\  \varepsilon_{ab} f_{\bar i, b}^{\dagger}   \end{pmatrix} \quad , \quad i=1,2,3,4
\end{eqnarray}
In Table \ref{table1}, we summarize the transformation rules of the spinor fields. 
\begin{table}[t]
\begin{spacing}{2}
\centering
\begin{tabular}{|c|c|c|c|c|} \hline
 & $T_{x,y}$   & $R_{\pi/2}$ & $I_{xy}$ & $\mathcal{T}$  \\
 \hline  \hline
$ ~\Psi_{1,a}~$ & $~\Psi_{1,a} e^{-i k_{1x,y} }~$  & $~i \tau^z \Psi_{3,a}~$  &  $~i \tau^z \Psi_{2,a}~$ & $- \tau^y \Psi_{1,a}$ \\
\hline 
$ ~\Psi_{2,a}~$ & $~\Psi_{2,a} e^{-i k_{2x,y} }~$  & $~i \tau^z \Psi_{4,a}~$  &  $~i \tau^z \Psi_{1,a}~$ & $- \tau^y \Psi_{2,a}$ \\
\hline 
$ ~\Psi_{3,a}~$ & $~\Psi_{3,a} e^{-i k_{3x,y} }~$ & $-i \tau^x \varepsilon_{ab} \tilde \Psi_{1,b}$  &  $-i \tau^x \varepsilon_{ab} \tilde \Psi_{4,b}$ & $- \tau^y \Psi_{3,a}$ \\
\hline 
$ ~\Psi_{4,a}~$ & $~\Psi_{4,a} e^{-i k_{4x,y} }~$  & $-i \tau^x \varepsilon_{ab} \tilde \Psi_{2,b}$  &  $-i \tau^x \varepsilon_{ab} \tilde \Psi_{3,b}$ & $- \tau^y \Psi_{4,a}$ \\
\hline
\end{tabular}
\end{spacing}
\caption{Symmetry transformations of the spinor fields under square
lattice symmetry operations.  
$T_{x,y}$: translation by one lattice
spacing along the $x,y$ direction; $R_{\pi/2}$: 90$^\circ$
rotation about a lattice site ($x\rightarrow y,y\rightarrow-x$); $I_{xy}$: reflection
about the $x=y$ axis ($x\rightarrow y,y\rightarrow x$);
$\mathcal{T}$: time-reversal, defined as a symmetry (similar to
parity) of the imaginary time path integral and the conjugate fields are transformed to $\Psi^{\dagger} \rightarrow \Psi^{\dagger} \tau_y$. 
Note that such a $\mathcal{T}$ operation is not anti-linear. Also, the notation, $\tilde \Psi_{i,b} =  (\Psi_{i,b}^{\dagger})^{T} $ is used for convenience. }
\label{table1}
\end{table}

Among various combinations of bilinear terms, the following operators are of interest. 
\begin{eqnarray}
O_{\Delta} & \equiv & +\Psi_{1,a}^{\dagger} \tau^x \Psi_{1,a} -  \Psi_{2,a}^{\dagger} \tau^x \Psi_{2,a} \nonumber \\
&& -\Psi_{3,a}^{\dagger} \tau^x \Psi_{3,a} +\Psi_{4,a}^{\dagger} \tau^x \Psi_{4,a}   \\
O_{\zeta} &\equiv&  +\Psi_{1,a}^{\dagger} \tau^x \Psi_{1,a} +  \Psi_{2,a}^{\dagger} \tau^x \Psi_{2,a} \nonumber \\
&& +\Psi_{3,a}^{\dagger} \tau^x \Psi_{3,a} +\Psi_{4,a}^{\dagger} \tau^x \Psi_{4,a}  \\
O_{\eta} &\equiv&  +\Psi_{1,a}^{\dagger} \tau^z \Psi_{1,a} -  \Psi_{2,a}^{\dagger} \tau^z \Psi_{2,a} \nonumber \\
&& -\Psi_{3,a}^{\dagger} \tau^z \Psi_{3,a} +\Psi_{4,a}^{\dagger} \tau^z \Psi_{4,a}  \\
O_{\rho} &\equiv&  +\Psi_{2,a}^{\dagger} \tau^z \Psi_{3,a} +  \Psi_{3,a}^{\dagger} \tau^z \Psi_{2,a} 
\end{eqnarray}
The operator, $O_\Delta$, is invariant under all the lattice symmetries.
This is just $d$ wave pairing term's low energy expression. 
Therefore, inserting the operator, $O_{\Delta}$, in the original Lagrangian is certainly allowed by the symmetry consideration. 
For the future convenience, let us write down the fermion Hamiltonian with pairing explicitly.
\begin{eqnarray}
H_{f} &=& \sum  \begin{pmatrix} \Psi^{\dagger}_{1,a} \\ \Psi^{\dagger}_{2,a} \end{pmatrix} \begin{pmatrix}   \varepsilon_1 \tau^z + \Delta \tau^x& 0\\ 0&   \varepsilon_2\tau^z - \Delta \tau^x\end{pmatrix} \begin{pmatrix} \Psi_{1,a} \\ \Psi_{2,a}  \end{pmatrix} \nonumber \\
&+& \sum  \begin{pmatrix} \Psi_{3,a}^{\dagger} \\ \Psi_{4,a}^{\dagger} \end{pmatrix} \begin{pmatrix}   \varepsilon_3\tau^z - \Delta \tau^x&0 \\0 &  \varepsilon_4 \tau^z + \Delta \tau^x\end{pmatrix} \begin{pmatrix} \Psi_{3,a} \\ \Psi_{4,a} \end{pmatrix} \nonumber \\
\label{hmatrix}
\end{eqnarray}
$\varepsilon_i(k) = \vec{v}_i \cdot \vec{k} + a~v_f^2 \vec{k}^2  $.
Note that $\Delta$ is a given constant here.

The $O_{\zeta,\eta}$ operators are also interesting, and they are transformed as follows: 
\begin{eqnarray}
T_{x,y} & : &  O_{\zeta,\eta} \rightarrow +O_{\zeta,\eta} \nonumber \\
R_{\pi/2} & : &  O_{\zeta,\eta} \rightarrow - O_{\zeta,\eta} \nonumber \\
I_{xy} & : &  O_{\zeta,\eta} \rightarrow -O_{\zeta,\eta} \nonumber \\
\mathcal{T} & : & O_{\zeta,\eta} \rightarrow +O_{\zeta,\eta}
\end{eqnarray}
Therefore, $O_{\zeta,\eta}$ operators have nematic ordering symmetries.
The difference between two operators are that $O_{\zeta}$ is from the pairing channel and $O_{\eta}$ is from the density channel. 

The final operator $O_{\rho}$ describes a charge density wave order parameter, which has horizontal ordering wavevector in this case.
Such ordering was considered a candidate of the ``pseudogap'' phase and the high energy $4 a_0$ ordering pattern in the 
cuprates \cite{kohsaka}.
Notice that the hot spots are in general not linked by $(\pi/4,0)$, 
but linked by the ordering vector, $\vec{Q}_{CDW} =\vec{k}_{f2}-\vec{k}_{f3}$, which is consistent with the observations
of the Hudson and collaborators \cite{hudson}.
For a thorough consideration of the charge density wave, we need to investigate ``hot-spots'' for the charge density wave order and start from the beginning. 
But because the calculations are identical in both cases, we consider the charge ordering within the present SDW theory. 

The original theory can be extended by introducing the other order parameters from the above microscopic consideration.
The total theory is 
\begin{eqnarray}
\mathcal{L}_{\varphi} &=&  \frac{1}{2}  (\partial_{\tau} \vec \varphi)^2 + \frac{1}{2}(\nabla \vec  \varphi)^2 + \frac{r}{2} ( \vec \varphi)^2 + \frac{u}{4} ( \vec \varphi^2)^2 \nonumber \\
\mathcal{L}_{\Psi} &=& \Psi^{\dagger}_{i,a} (\partial_{\tau} + \mathcal{H}_0) \Psi_{i,a} \nonumber \\
\mathcal{L}_{\eta} &=& \frac{1}{2} (\partial \eta)^2 + s \eta^2 + \cdots \nonumber \\
\mathcal{L}_{\rho} &=& \frac{1}{2} (\partial \rho)^2 + w \rho^2 + \cdots \nonumber \\
\mathcal{L}_{\varphi - \Psi} &=& \lambda_{\varphi} \vec{\varphi} \cdot \left[\Psi^{\dagger}_{1,a}  \vec{\sigma}_{ab} \tau^x \varepsilon_{bc}  \tilde \Psi_{2,c} + \Psi^{\dagger}_{3,a}  \vec{\sigma}_{ab} \tau^x \varepsilon_{bc}  \tilde \Psi_{4,c} \right] \nonumber \\
\mathcal{L}_{\eta-\Psi} &=& \lambda_{\eta} \eta \cdot  O_{\eta} =  \lambda_{\eta} \eta \sum_{j} m_j \Psi_{j,a}^{\dagger} \tau^z \Psi_{j,a}  \nonumber \\
\mathcal{L}_{\rho-\Psi} &=& \lambda_{\rho} \rho \cdot  O_{\rho} =  \lambda_{\rho} \rho (\Psi_{2,a}^{\dagger} \tau^z \Psi_{3,a} +\Psi_{3,a}^{\dagger} \tau^z \Psi_{2,a})  \nonumber \\
\mathcal{L'}_{\zeta} &=& \frac{1}{2} (\partial \zeta)^2 + s_1 \zeta^2 + \cdots \nonumber \\
\mathcal{L'}_{\zeta-\Psi} &=& \lambda_{\zeta} \zeta \cdot  O_{\zeta} = \lambda_{\zeta} \zeta \cdot  \sum_j  \Psi_{j,a}^{\dagger} \tau^x \Psi_{j,a}, \label{eq:total}
\end{eqnarray} 
where $\mathcal{H}_0$ is the spatial representation of the Eq. (\ref{hmatrix}) and $m_{1,4}=1, m_{2,3}=-1$ . 
Note that the $\mathcal{L'}$ terms are higher order in terms of the pairings.

Before going further, let us remark the meaning of the above consideration. 
In this extended theory, we have found the two different nematic order channels. 
One order parameter, $\eta$, couples to the density of fermions. 
The other one, $\zeta$, couples to the pairing, which suggests they have different charges.
Many previous works focus on the pairing channel nematic order with the nodal fermions of the $d_{x^2 -y^2}$ superconductivity. \cite{vojta1,vojta2,vicari1,vicari2,kim,huh}
The nematic order naturally induces different pairings, $\Delta_x \neq \Delta_y$, which corresponds to the condensation of $\zeta$ in this case. 
The original nodal fermions are gapped and the nodes become shifted depending on the sign of the gap function.
Within the nodal fermion theory the density channel is not allowed by the square lattice symmetry. 
In our spin-fermion model, the density channel is surely allowed and can see the effect under the small superconductivity.
Note that the pairing channel is higher order in the pairing amplitude $\Delta$, which we will assume it is small here.

\section{Critical point shifts} \label{QCPshift}
\begin{figure}
\includegraphics[width=2.0 in]{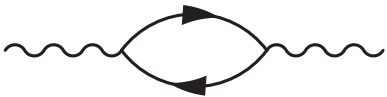}
\caption{Susceptibility Feynman diagram. The wavy line is for the order parameters like the SDW or the nematic order. The vertex matrix elements depend on the order parameters. }
\label{diagrams}
\end{figure}

In this section, to study influence of the superconductivity on quantum critical points in general, we come back to general cases, 
including the incommensurate cases for the cuprates. 
We can easily generalize the previous discussion to the incommensurate case:
the changes are mainly in the fermions' spectra, which have different velocities and gap functions between two SDW linked points, 
and the existence of the complex two vector fields for the SDW. 
We will see effects of superconductivity on the quantum critical points by evaluating the diagrams in Fig.~\ref{diagrams} for various order parameters with and without superconductivity. 

\subsection{Spin density wave}
To investigate the effect of the superconductivity on the SDW critical point, let us focus on the SDW order near the criticality with and without superconductivity. 
The interaction between the SDW order and fermions naturally affects the critical point as we saw in the Eq. (\ref{Hertz}).
If we allow the superconductivity, the main effect of the superconductivity is gapping out the fermion surfaces, so the contribution to the critical point has to be changed.
The amount of the change can be obtained by evaluating the susceptibility with and without the pairing.
In each case, the critical point is affected with the loop contribution and we can evaluate them with the Hamiltonian, Eq. (\ref{hmatrix}).
The total susceptibility for the one ordering wavevector,say $1$, is 
\begin{eqnarray}
\chi^{\phi}_{\Delta} &=& \sum_{k_i - k_j = \vec{Q_1}} \chi^{\phi}_{\Delta,ij} \nonumber
\end{eqnarray}
Due to the symmetry it is enough to choose one of the linking hot-spots. 
 For example, the SDW fluctuation between $k_{f,1}$ and $k_{g,\bar 2}$ is 
\begin{eqnarray}
&&\chi^{\phi}_{\Delta,1\bar{2}} \nonumber \\
&&=  (-) 2 \lambda_{\varphi}^2 \int_{k,\omega} \frac{i \omega + \varepsilon_1(k)}{\omega^2+\varepsilon_1^2+\Delta_+^2} \frac{i \omega + \varepsilon_{\bar{2}}(k)}{\omega^2+\varepsilon_2^2+\Delta_-^2} \nonumber \\
&&+ 2 \lambda_{\varphi}^2  \int_{k,\omega}  \frac{\Delta_+}{\omega^2+\varepsilon_1^2+\Delta_+^2}  \frac{\Delta_-}{\omega^2+\varepsilon_2^2+\Delta_-^2} \nonumber \\
&&= 2 \lambda_{\varphi}^2\int_{k,\omega} \frac{ \omega^2 - \varepsilon_1(k)\varepsilon_{\bar{2}}(k)+\Delta_+ \Delta_-}{(\omega^2+\varepsilon_1^2+\Delta_+^2)(\omega^2+\varepsilon_2^2+\Delta_-^2)} . \label{ss1}
 \end{eqnarray}
The two gap functions arise from the two bands in pnictides, and from the distance differences from the nodal point in the cuprates.
To parametrize gap difference, let us introduce one parameter, $\alpha$ as $\Delta_{\pm} = \Delta(1\pm \alpha)$;
thus $\alpha$ characterizes the distinction between the two hot spots, which become crystallographically equivalent 
$\alpha =0$. For the cases under consideration here, $\alpha=0$ for a commensurate $(\pi, \pi)$ wavevector
for the SDW ordering for the cuprate case, while $\alpha \neq 0$ in all other cases.
As we showed in the Eq. (\ref{Hertz}), the susceptibility function contributing to the critical point and the relative critical point shifts with and without superconductivity is defined as  
\begin{eqnarray}
r_c^{\Delta} &=& \chi^{\phi}_{\Delta} = \chi^{\phi}_{0} - (\chi^{\phi}_{0} - \chi^{\phi}_{\Delta}) = r_c^0 - \delta_{r}.
\end{eqnarray}
A positive $\delta_r$ implies the critical point shifts to shrink the SDW region leading to competition between the SDW and the superconductivity, 
while the negative sign implies attraction between the SDW and the superconductivity. 

\begin{figure}
\includegraphics[width=3.0 in]{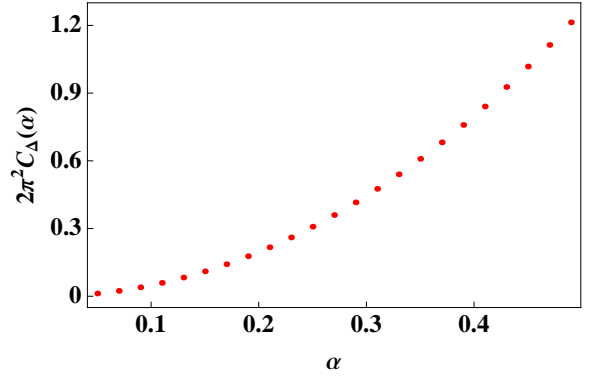}
\caption{ The SDW critical point shift with the relative gap parameter $\alpha$, $C_{\Delta}(\alpha)$.  }
\label{shift}
\end{figure}

As we can see, the fermion loop formula, $\chi^{\phi}_{\Delta}$, is divergent in the ultraviolet limit without the curvature term.
But the critical point shift, which was defined as difference between different pairing magnitudes, is well-defined. 
This is because the susceptibility function contains two independent momentum components unless the two Fermi velocities are parallel, which requires hot spots are the same as the nodes. 
We exclude such special case in this paper. 
Therefore, the curvature term, $a$, can be dropped, 
and we assume  $|\Delta| a \ll 1$. 
So, the dispersion relation $\varepsilon_i(k) = \vec{v_i} \cdot \vec{k}$ will be used for evaluation. 
\begin{eqnarray}
\delta_r &=& \chi^{\phi}_{0} - \chi^{\phi}_{\Delta}  \nonumber \\
&=&N_f \lambda_{\varphi}^2\int_{k} \Biggl[  \frac{1}{|v_1 \cdot k| + |v_2 \cdot k|} \Biggl(1+ \frac{v_1 \cdot k~ v_2 \cdot k}{|v_1 \cdot k| |v_2 \cdot k|} \Biggr) \nonumber \\
&& - \frac{1}{\sqrt{(v_1 \cdot k)^2 + \Delta_{+}^2}+\sqrt{(v_2 \cdot k)^2 + \Delta_{-}^2}} \nonumber \\
&& \times \Biggl(1+ \frac{v_1 \cdot k ~v_2 \cdot k}{\sqrt{(v_1 \cdot k)^2 + \Delta_{+}^2}\sqrt{(v_2 \cdot k)^2 + \Delta_{-}^2}}\nonumber \\
&& +  \frac{\Delta_{+} \Delta_{-}}{\sqrt{(v_1 \cdot k)^2 + \Delta_{+}^2}\sqrt{(v_2 \cdot k)^2 + \Delta_{-}^2}}\Biggr)\Biggr]  \nonumber \\
&=& N_f ~ \lambda_{\varphi}^2 \frac{C_\Delta (\alpha)}{|\sin(\theta_1-\theta_2)|} ~\frac{|\Delta|}{v_{f1} v_{f2}},
\end{eqnarray} 
where the Fermi velocities are defined as
\begin{eqnarray}
&&\vec{v_1} = v_{f1} (\cos(\theta_1), \sin(\theta_1)) ~ , ~ \vec{v_2} = v_{f2}(\cos(\theta_2), \sin(\theta_2) \nonumber
\end{eqnarray}
Therefore, the angle dependence and the relative gap function determine the critical point shift. 
The angle dependence on the two Fermi velocities indicates 
that  more parallel velocities implies a larger critical point shift. 
So the perpendicular Fermi velocities of SDW participating fermions have the smallest critical point shift.
As we can see, the sine function dependence indicates that the collinear Fermi velocities are dangerous in our calculation.
In this case, we need to keep $a$ from the start, and the functional behavior becomes the same as the nematic case. 
The coefficient, $C(\alpha)$, is 
\begin{eqnarray}
&&C_{\Delta} (\alpha) = \frac{1}{4 \pi^2} \int d q_x d q_y \Biggl[ \nonumber \\
&&\frac{1}{|q_x|+|q_y|}- \frac{1}{\sqrt{q_x^2 + (1-\alpha)^2}+\sqrt{q_y^2 + (1+\alpha)^2}} \nonumber \\
&& \times \Biggl(1+ \frac{1 - \alpha^2}{\sqrt{q_x^2 + (1-\alpha)^2}\sqrt{q_y^2 + (1+\alpha)^2}} \Biggr) \Biggr]. \label{resC}
\end{eqnarray}
As pointed out in Section~\ref{intro}, the sign of $C_\Delta (\alpha)$ is not immediately evident from 
Eq.~(\ref{resC}): the small $\vec{q}$ region near the hot spot contributes a positive sign, while that
from large $\vec{q}$ has a negative sign. A numerical evaluation of Eq.~(\ref{resC}) shows that the result is
indeed non-negative for all $\alpha$, and the result is shown in 
Fig.~\ref{shift}: $C_\Delta ( \alpha)$ increases monotonically with increasing $\alpha$.

A curious feature of Fig.~\ref{shift} is that $C_\Delta ( 0)  = 0$: it vanishes for the case of equivalent hot spots.
However, it is not at all evident from the integral in Eq.~(\ref{resC}) that it equals zero for $\alpha =0$.
This is more easily proved from the original expression in Eq.~(\ref{ss1}), by reversing the order of frequency
and momentum integration. Evaluating first the momentum integration in Eq.~(\ref{ss1}) (after linearizing the dispersion
about the Fermi surfaces) we find
\begin{eqnarray}
\chi^{\phi}_{\Delta,1\bar{2}} &=&  \lambda_{\varphi}^2 \frac{1}{2v_{f1} v_{f2} |\sin(\theta_1-\theta_2)|} \nonumber \\
&~&~~\times
\int_\omega \frac{ \omega^2 + \Delta_+ \Delta_-}{\sqrt{\omega^2 + \Delta_+^2} \sqrt{\omega^2 + \Delta_-^2}}
\end{eqnarray}
It is now evident that the critical point shift vanishes for $\Delta_+ = \Delta_-$, for then the above result
becomes independent of the value of the gap.

If we consider the state slightly off the criticality. {\em e.g.\/} by considering finite temperature, then the low energy physics is governed by the temperature. 
In such a case, the shift becomes an analytic function of the superconducting gap, which means quadratic gap functions scaled with another energy scale, instead of the linear gap. 
Therefore, we can understand the non-analytic behavior of the linear gap as a property of the quantum critical points.   

\subsection{Charge density wave}
The CDW ordering operator,$O_{\rho}$, with a specific ordering wave vector, $\vec{Q}_{CDW} =\vec{k}_{f2}-\vec{k}_{f3}$,  was introduced in the previous section. 
The CDW has different characteristics from the SDW. For example, instead of the spin dependent vertex, it has density type operators linking two hot spots, and also the linked hot-spots' gap functions have the same pairing sign and magnitude. 
Combining all of these, the susceptibility for the CDW with superconductivity is 
\begin{eqnarray}
&&\chi^{\rho}_{\Delta, 2 3} \nonumber \\
&&=  (-) 2 \lambda_{\rho}^2 \int_{k,\omega} \frac{i \omega + \varepsilon_3(k)}{\omega^2+\varepsilon_3^2+\Delta_{+}^2} \frac{i \omega + \varepsilon_{2}(k)}{\omega^2+\varepsilon_2^2+\Delta_{+}^2} \nonumber \\
&&+ 2 \lambda_{\rho}^2  \int_{k,\omega}  \frac{\Delta_{+}}{\omega^2+\varepsilon_3^2+\Delta_{+}^2}  \frac{\Delta_{+}}{\omega^2+\varepsilon_2^2+\Delta_{+}^2} \nonumber \\
&&= 2 \lambda_{\rho}^2\int_{k,\omega} \frac{ \omega^2 - \varepsilon_3(k)\varepsilon_{2}(k)+\Delta_{+} \Delta_{+}}{(\omega^2+\varepsilon_3^2+\Delta_{+}^2)(\omega^2+\varepsilon_2^2+\Delta_{+}^2)} .
 \end{eqnarray}
The CDW critical point shift is
 \begin{eqnarray}
\delta_w &=& \chi^{\rho}_{0} - \chi^{\rho}_{\Delta}  \nonumber \\
&=&\frac{N_f}{2} \lambda_{\rho}^2\int_{k}  \Biggl[ \frac{1}{|v_3 \cdot k| + |v_2 \cdot k|} 
\Biggl(1- \frac{v_3 \cdot k~ v_2 \cdot k}{|v_3 \cdot k| |v_2 \cdot k|} \Biggr) \nonumber \\
&& - \frac{1}{\sqrt{(v_3 \cdot k)^2 + \Delta_{+}^2}+\sqrt{(v_2 \cdot k)^2 + \Delta_{+}^2}} \nonumber \\
&& \times \Biggl (1- \frac{v_3 \cdot k ~v_2 \cdot k}{\sqrt{(v_3 \cdot k)^2 + \Delta_{+}^2}\sqrt{(v_2 \cdot k)^2 + \Delta_{+}^2}}\nonumber \\
&& +  \frac{\Delta_{+} \Delta_{+}}{\sqrt{(v_3 \cdot k)^2 + \Delta_{+}^2}\sqrt{(v_2 \cdot k)^2 + \Delta_{+}^2}}
\Biggr) \Biggr]  \nonumber \\
&=& \frac{N_f}{2} ~ \lambda_{\rho}^2 \frac{C_\Delta (0)}{|\sin(\theta_2-\theta_3)|} ~\frac{|\Delta_+|}{v_{f2}^2}, \label{cdwshift}
\end{eqnarray} 
where the Fermi velocities are defined as
\begin{eqnarray}
&&\vec{v_2} = v_{f2} (\cos(\theta_2), \sin(\theta_2)) ~ , ~ \vec{v_3} = v_{f2}(\cos(\theta_3), \sin(\theta_3), \nonumber
\end{eqnarray}
where the angle between $\vec{v}_{2,3}$ is ($\theta_2- \theta_3$), and the velocities are not collinear in general. 
Interestingly, the calculation itself is quite similar to the SDW's formally.
However, a key difference is that the participating fermions have the exactly same gap functions and Fermi velocities for the CDW, and this is guaranteed by symmetry. Consequently, the prefactor in Eq.~(\ref{cdwshift}) is $C_\Delta (\alpha=0)$
which is zero. Thus the shift in the CDW critical point vanishes at this order.

We can estimates higher order in $\Delta$ by evaluating Eq.~(\ref{cdwshift}) with a finite momentum cutoff $\Lambda$.
This introduces dependence of the result on $\Delta/\Lambda$, which is evaluated in Appendix~\ref{app:cdw}.
We find a net competing effect, but this is formally higher order in $\Delta $ and so parametrically 
smaller than the SDW case.

\subsection{Nematic order}
With the extension of the spin-fermion theory, we can consider the nematic order parameter within the theory. 
As we mentioned before, there are two channels for the nematic order, but for the case of a small pairing gap, 
it is enough to consider the density channels. 
The critical point correction can be evaluated from the fermion loop calculation as before, and it is 
\begin{eqnarray}
\frac{1}{N_f} \chi^{\eta}_{\Delta} 
 &=& (-1)\lambda_{\eta}^2 \int_{k,\omega} {\rm Tr}\Biggl(\tau^z \frac{i \omega + \varepsilon_k \tau^z + \Delta_i \tau^x }{\omega^2 + \varepsilon_k^2 + \Delta_i^2} \nonumber \\
 && ~~~~~~~~~~~~\times \tau^z \frac{i \omega + \varepsilon_k \tau^z + \Delta_i \tau^x }{\omega^2 + \varepsilon_k^2 + \Delta_i^2} \Biggr) \nonumber \\
&=& 2 \lambda_{\eta}^2  \int_{k,\omega} \Biggl( \frac{1}{\omega^2 + \varepsilon_k^2 + \Delta_i^2} -\frac{2 \varepsilon_k^2}{(\omega^2 + \varepsilon_k^2 + \Delta_i^2)^2} \Biggr) \nonumber \\
&=&  \lambda_{\eta}^2  \int_k \frac{\Delta_i^2}{(\varepsilon_k^2+\Delta_i^2)^{3/2}}  
\end{eqnarray}
Note that there is a crucial difference in this integration compared to the previous critical point shifts. 
Because the nematic order parameter consists of particle and hole with same Fermi velocity, the pathology of collinear dispersions are always present in the nematic phase transition. 
Therefore, we need to keep the curvature term, $a$, and then we have the susceptibility as
\begin{eqnarray}
\frac{1}{N_f} \chi^{\eta}_{\Delta} 
 &=& 2 \lambda_{\eta}^2  \int_{k,\omega} \Biggl( \frac{1}{\omega^2 + \varepsilon_k^2 + \Delta_i^2} -\frac{2 \varepsilon_k^2}{(\omega^2 + \varepsilon_k^2 + \Delta_i^2)^2} \Biggr)  \nonumber \\
&=&  \lambda_{\eta}^2  \int_k \frac{\Delta_i^2}{(\varepsilon_k^2+\Delta_i^2)^{3/2}}  \nonumber \\
&=& \lambda_{\eta}^2  D_0 \int^{\infty}_{-1/(4a)} d \varepsilon \frac{\Delta_i^2}{(\varepsilon^2+\Delta_i^2)^{3/2}} \nonumber \\
&=&   \lambda_{\eta}^2 \frac{1}{4 \pi}\frac{1}{v_f^2} \frac{1}{a} \left( 1- 8 (|\Delta_i| a)^2 \right) ,
\end{eqnarray}
where $D_0$ is the constant density of state. The lower cut-off is from the dispersion relation $\varepsilon_k =  \vec{v} \cdot \vec{k} + a v^2 k^2 $.
Note that this integration is well-defined in both ultra-violet and infra-red regions.
Even in the SDW the colinear Fermi velocity hot spots whose cases are excluded in this paper also suffer similar problems. 

The critical point shift of the nematic ordering, then, is 
\begin{eqnarray}
\delta_{c}^{\eta} \equiv \chi^{\eta}_{0} - \chi^{\eta}_{\Delta} =  N_f \lambda_{\eta}^2  \frac{C_{\eta}}{v_f^2}  \frac{1}{a} (|\Delta_i| a)^2 ,
\end{eqnarray}
where $C_{\eta} = \frac{2}{\pi} $.
Note that the nematic critical point shift does not contain the linear gap behavior like the SDW. 
Instead it starts from the second order and it is analytic in terms of the gap function. 
Such a term describes usual competing term of the Landau-Ginzburg theory, as was discussed in 
Section~\ref{intro}.
Parametrically the nematic ordering is more stable than the spin density wave under the ``weak'' superconductivity. 

\section{Conclusions} \label{conclusion}

\begin{figure}
\includegraphics[width=3in]{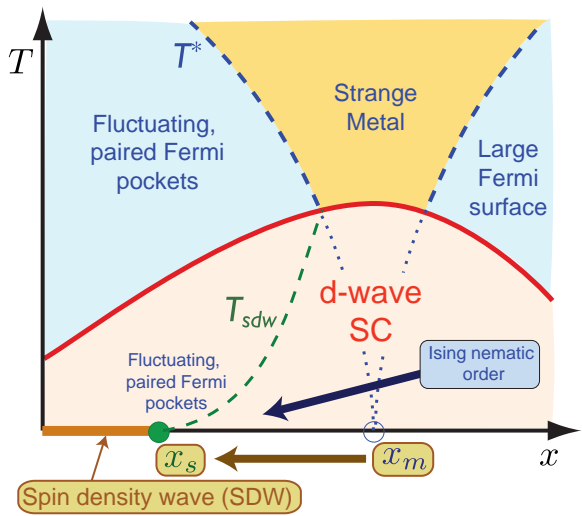}
\caption{Cuprate phase diagram adapted 
from Ref.~\onlinecite{eg,qcnp}. Here $x$ is hole doping, $x_m$ is the position
of the SDW critical point in the metal, and $x_s$ is the SDW critical point in the insulator. The shift between
$x_m$ and $x_s$ represents the consequence of Eq.~(\ref{sdwshift}). The regions with 
`fluctuating Fermi pockets' have renormalized classical thermal fluctuations of SDW order. Ising-nematic 
ordering is expected for $T < T^\ast$, and consistent with Fig.~\ref{totalshift}, this regime
is not sensitive to the onset of superconductivity.}
\label{figcuprates}
\end{figure}

Our main results, summarized in Fig.~\ref{totalshift}, have a natural application to the physics of the cuprates.
For SDW ordering, we have shown that there is a large shift in the quantum critical point due
to the onset of superconductivity, represented by Eq.~(\ref{sdwshift}). In contrast, for the Ising-nematic order,
when considered as an independent order parameter, there is a significantly smaller shift, of order
that expected in Landau theory in Eq.~(\ref{landaushift}). These results provide a natural
basis for the phase diagram proposed in earlier work \cite{eg,qcnp}, which we reproduce here
in Fig.~\ref{figcuprates}.
The large shift in the SDW ordering between the metal and the superconductor is represented by 
the arrow from $x_m$ to $x_s$. We assume that the Ising-nematic ordering has an onset around $x_m$,
and this is barely shifted by the onset of superconductivity, as implied by Fig.~\ref{totalshift}. 
Consequently, long-range Ising-nematic ordering can survive for $x> x_s$, as is indicated in 
Fig.~\ref{figcuprates}. These results are consistent with recent observations of Ising-nematic 
ordering \cite{hinkov08a,taill10b,lawler10} in the hole-doped cuprates.

In an applied magnetic field, as was discussed in Refs.~\onlinecite{qcnp,zhang}, the point $x_s$ eventually
merges with $x_m$, so that the SDW transition in the high-field normal state takes place at $x=x_m$.
Given this, we expect that the Ising nematic transition will also merge (or become very close to)
with the SDW transition at high fields.

A notable feature of Fig.~\ref{figcuprates} is the ``back-bending'' of the crossover line bounding
the region where there are `renormalized classical' fluctuations of local SDW order: this region
is bounded by the line labeled $T^\ast$ in the normal state, and by the line labeled $T_{\rm sdw}$
in the superconducting state. As we have argued \cite{eg,qcnp}, this is a natural consequence
of the shift of the in SDW quantum critical point from $x_m$ to $x_s$.

\begin{figure}
\includegraphics[width=3in]{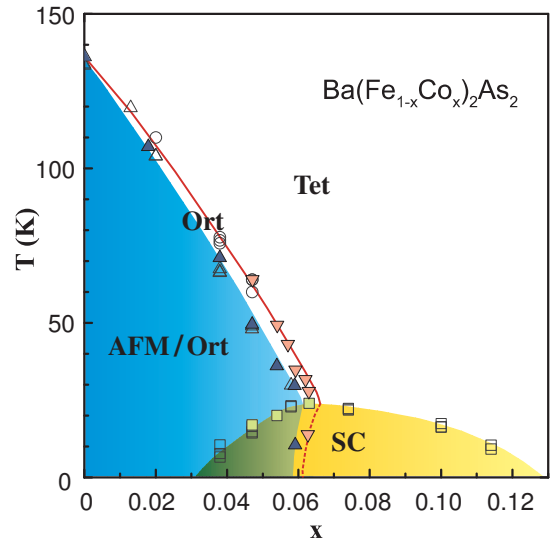}
\caption{Phase diagram for Ba[Fe$_{1-x}$Co$_x$]$_2$As$_2$
from Refs.~\onlinecite{joerg1,joerg2}.  The back-bending of the
SDW ordering transition in the superconducting phase is similar to that of $T_{sdw}$ in Fig.~\ref{figcuprates}.
Here, rather than renormalized classical SDW fluctuations, we have true long-range order
indicated by `AFM'. The Ising-nematic order is present in the phase labeled
`Ort' and absent in that labeled `Tet'. Note that, unlike the cuprates, 
the Ising-nematic transition follows the SDW ordering
transition in the superconducting state too.
}
\label{figpnictides}
\end{figure}

Turning to the pnictides, we note that the back-bending of the SDW ordering has been clearly
seen in recent experiments, as shown in Fig.~\ref{figpnictides}.
Here, because the stronger 3-dimensionality of the crystal structure and the commensurate
wavevector, the region of renormalized classical SDW fluctuations becomes a region
of true long-range order, and so is more easily detected by
neutron scattering. However, Fig.~\ref{figpnictides} differs from the phase diagram in Fig.~\ref{figcuprates}
in one important aspect: note that the Ising-nematic transition in Fig.~\ref{figpnictides} closely tracks
the SDW transition in both the normal and superconducting states, rather than separating from it in the 
superconducting state as in Fig.~\ref{figcuprates}. This means that the shift in the nematic ordering
transition due to superconductivity is not significantly smaller than that of the SDW transition. 
This is in conflict with the situation outlined in Fig.~\ref{totalshift}, where the nematic transition
hardly shifts relative to the SDW transition.

This difference between our computations and the pnictide phase diagram in Fig.~\ref{figpnictides}
implies that the Ising-nematic transition in the pnictides is not an independent instability
associated with the electrons near the Fermi surface. 
For if it were, our computations show that it would barely notice the onset of superconductivity. We now argue that the phase diagram in Fig.~\ref{figpnictides} can be understood
if we assume that the Ising-nematic ordering is primarily induced by its coupling to the 
square of the SDW order \cite{kivelson,cenke}. Thus, in addition to the coupling of the nematic order, $\eta$,
to the fermions in Eq.~(\ref{eq:total}), we need to add its coupling to the SDW order:
\begin{equation}
\mathcal{L}_{\eta - \varphi} = \widetilde{\lambda} \eta \left( \vec{\varphi}_x^2 - \vec{\varphi}_y^2 \right),
\end{equation}
where $\vec{\varphi}_x$ ($\vec{\varphi}_{y}$) is the SDW ordering at wavevector $(\pi, 0)$ ($(0,\pi)$).
Then a correction of order $|\Delta|$ to the SDW fluctuations from the onset of superconductivity,
will feed into a similar correction to the Ising-nematic fluctuations via a perturbation theory in $\widetilde{\lambda}$.
Thus our conclusion is that $\widetilde{\lambda}$ is the dominant coupling which induces Ising-nematic 
order in the pnictides. A similar conclusion 
appears to have been reached recently by Kimber {\em et al.} \cite{mazin} based upon their analysis
of the STM observations of Chuang {\em et al.} \cite{chuang}. 
In contrast, for the cuprates, the influence of $\widetilde{\lambda}$ appears significantly weaker.

\subsubsection*{Acknowledgements}
We thank E.~Berg, A.~Chubukov, L.~Fu, T.~Imai, M.~Metlitski, J.~Schmalian, L.~Taillefer, and C. Xu for useful discussions. 
We are grateful to A.~I.~Goldman and J.~Schmalian for permission to reproduce Fig.~\ref{figpnictides}.
This research was supported by the National Science Foundation under grant DMR-0757145, by the FQXi
foundation, and by a MURI grant from AFOSR. 
E. G. Moon was also supported by the Samsung Scholarship.

\appendix

\section{Symmetry}
\label{app:symmetry}

\begin{table}[t]
\begin{spacing}{2}
\centering
\begin{tabular}{|c|c|c|c|c|} \hline
 & $T_x$ & $T_y$ & $R_{\pi/2}$ & $I_{xy}$  \\
 \hline  \hline
$ ~f_{1,a}~$ & $~f_{1,a} e^{-i k_{1x} }~$ & $ ~f_{1,a} e^{-i k_{1y} }~$ & $~i f_{3,a}~$  &  $~i f_{2,a}~$  \\
\hline 
$ ~f_{2,a}~$ & $~f_{2,a} e^{-i k_{2x} }~$ & $ ~f_{2,a} e^{-i k_{2y} }~$ & $~i f_{4,a}~$  &  $~i f_{1,a}~$  \\
\hline 
$ ~f_{3,a}~$ & $~f_{3,a} e^{-i k_{3x} }~$ & $ ~f_{3,a} e^{-i k_{3y} }~$ & $~i f_{\bar 1,a}~$  &  $~i f_{\bar 4,a}~$  \\
\hline 
$ ~f_{4,a}~$ & $~f_{4,a} e^{-i k_{4x} }~$ & $ ~f_{4,a} e^{-i k_{4y} }~$ & $~i f_{\bar 2,a}~$  &  $~i f_{\bar 3,a}~$  \\
\hline
$ ~f_{\bar 1,a}~$ & $~f_{\bar1,a} e^{i k_{1x} }~$ & $ ~f_{\bar 1,a} e^{i k_{1y} }~$ & $~i f_{\bar 3,a}~$  &  $~i f_{\bar 2,a}~$  \\
\hline 
$ ~f_{\bar 2,a}~$ & $~f_{\bar 2,a} e^{i k_{2x} }~$ & $ ~f_{\bar 2,a} e^{i k_{2y} }~$ & $~i f_{\bar 4,a}~$  &  $~i f_{\bar 1,a}~$  \\
\hline 
$ ~f_{\bar 3,a}~$ & $~f_{\bar 3,a} e^{i k_{3x} }~$ & $ ~f_{\bar 3,a} e^{i k_{3y} }~$ & $~i f_{1,a}~$  &  $~i f_{4,a}~$ \\
\hline 
$ ~f_{\bar 4,a}~$ & $~f_{\bar 4,a} e^{i k_{4x} }~$ & $ ~f_{\bar 4,a} e^{i k_{4y} }~$ & $~i f_{2,a}~$  &  $~i f_{3,a}~$  \\
\hline
\end{tabular}
\end{spacing}
\caption{Symmetry transformations of the hot-spot fields under square
lattice symmetry operations.} 
\label{spottable}
\end{table}

In this section, we set the notation for the symmetry transformation of the square lattice. 
Mean field Hamiltonian with the $d_{x^2 -y^2}$ needs the pairing term as  $\Delta_{i,i \pm x} =-\Delta_{i,i \pm y}$. 
\begin{eqnarray}
H_{MF} &=& - \sum_{<ij>} t_{ij} (c_{i a}^{\dagger} c_{j a} + h.c ) - \mu \sum_{i} c_{i,a}^{\dagger} c_{i, a}\nonumber \\
&& + \sum_{i,j} \Delta_{i,j}^{*} (\epsilon^{a b} c_{i a} c_{j b}) + h.c.
\label{hamiltonian}
\end{eqnarray}
The first line describes usual hopping terms on the square lattice, which gives the Fermi surface. 
We exclude special `nesting' type Fermi surfaces and assume there are points linked by the spin density wave ordering vector.

We start with lattice field transformations. 
\begin{eqnarray}
T_{x,y} & : & c_{a} (x) \rightarrow c'_{a} (x') =  c_{a}(x) \nonumber \\
R_{\pi/2} & : &  c_{a} (x) \rightarrow c'_{a} (x') = i c_{a}(x)\nonumber \\
I_{xy} & : & c_{a} (x) \rightarrow  c'_{a} (x') = i c_{a}(x) 
\end{eqnarray}
The rotation and reflection transformation attaches the factor $i$, which makes the $d$ wave pairing term invariant.
After writing the lattice fields with continuum field, we can obtain the transformation in Table. \ref{spottable}.
It is worthwhile to mention that at low energy or long-wavelength scale the fields at hot spots can be treated as independent fields.

Time reversal symmetry is obtained with low energy fields instead of the lattice fields, so we do not consider it here.
See the caption of the Table. \ref{table1}.
\begin{eqnarray}
\Psi_{i,a} \equiv \begin{pmatrix} f_{i,a} \\  \varepsilon_{ab} f_{\bar i, b}^{\dagger}   \end{pmatrix} \quad , \quad i=1,2,3,4
\end{eqnarray}
After introducing the Nambu spinors, bilinear spinors' transformations can be done easily.

Physical quantities are described with bilinear terms such as density and pairing interactions. 
Below several important bilinear terms are listed up to constants. 
\begin{eqnarray}
 \Psi_{1,a}^{\dagger} \tau^0 \Psi_{1,a} &=& f^{\dagger}_{1a} f_{1a} -  f^{\dagger}_{\bar1a} f_{\bar1a} \nonumber \\
 \Psi_{1,a}^{\dagger} \tau^z \Psi_{1,a} &=& f^{\dagger}_{1a} f_{1a} +  f^{\dagger}_{\bar1a} f_{\bar1a} \nonumber \\
 \Psi_{1,a}^{\dagger} \tau^x \Psi_{1,a} &=& \varepsilon_{ab}(f^{\dagger}_{1a} f^{\dagger}_{\bar1b} +  f_{\bar1b} f_{1a} )\nonumber \\
 \Psi_{1,a}^{\dagger} \tau^y \Psi_{1,a} &=&(-i) \varepsilon_{ab}(f^{\dagger}_{1a} f^{\dagger}_{\bar1b} -  f_{\bar1b} f_{1a} )\nonumber \\
 \Psi_{1,a}^{\dagger} \tau^x  \vec{\sigma}_{ab} \varepsilon_{bc} \tilde \Psi_{2,c} &=&-  (f^{\dagger}_{1a} \vec{\sigma}_{ab} f_{\bar2 b} +  f^{\dagger}_{2a} \vec{\sigma}_{ab} f_{\bar1 b} ) \nonumber \\
 \Psi_{1,a}^{\dagger} \tau^y  \vec{\sigma}_{ab} \varepsilon_{bc} \tilde \Psi_{2,c} &=& i( f^{\dagger}_{1a}  \vec{\sigma}_{ab}f_{\bar2 b} -  f^{\dagger}_{2a} \vec{\sigma}_{ab} f_{\bar1 b} ) \nonumber \\
  \Psi_{1,a}^{\dagger} \tau^0 \vec{\sigma}_{ab}\Psi_{1,b} &=& f^{\dagger}_{1a}  \vec{\sigma}_{ab} f_{1b} +  f^{\dagger}_{\bar1a}  \vec{\sigma}_{ab} f_{\bar1b} \nonumber \\
 \Psi_{1,a}^{\dagger} \tau^z  \vec{\sigma}_{ab} \Psi_{1,b} &=& f^{\dagger}_{1a}  \vec{\sigma}_{ab} f_{1b} -  f^{\dagger}_{\bar1a}  \vec{\sigma}_{ab} f_{\bar1b} \nonumber \\
 \Psi_{1,a}^{\dagger} \tau^x  \vec{\sigma}_{ab} \Psi_{1,b} &=& (f^{\dagger}_{1a} f^{\dagger}_{\bar1c} \varepsilon_{bc} +  f_{\bar1d} f_{1b} \varepsilon_{ad} ) \vec{\sigma}_{ab}\nonumber \\
 \Psi_{1,a}^{\dagger} \tau^y  \vec{\sigma}_{ab} \Psi_{1,b} &=&(-i)  (f^{\dagger}_{1a} f^{\dagger}_{\bar1c} \varepsilon_{bc} -  f_{\bar1d} f_{1b} \varepsilon_{ad} ) \vec{\sigma}_{ab}\nonumber \\
\end{eqnarray}

\section{The CDW critical point shift}
\label{app:cdw}

\begin{figure}
\includegraphics[width=3.0 in]{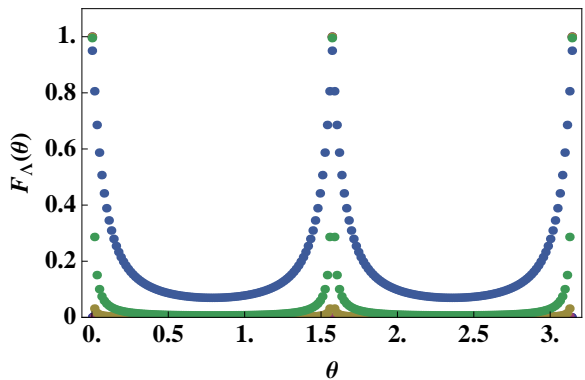}
\caption{ The CDW critical point shift varying the cutoff. The CDW has participating fermions which have the same gap functions, so it is $\alpha=0$ case formally.   The blue, green, dark-yellow correspond to the cutoff $10$, $10^2$, $10^3$. As we can see, the integrand is positive in any cases, but the numbers are getting smaller with larger cutoffs. }
\label{F}
\end{figure}

As we saw in the above, the critical point shift for the CDW is determined by the function,
\begin{eqnarray}
&&C_{\Delta} (0) = \frac{1}{4 \pi^2} \int d q_x d q_y \Biggl[ \frac{1}{|q_x|+|q_y|}\nonumber \\
&&- \frac{1}{\sqrt{q_x^2 + 1}+\sqrt{q_y^2 + 1}} \times \Biggl(
1+ \frac{1}{\sqrt{q_x^2 +1}\sqrt{q_y^2 + 1}}\Biggr) \Biggr] \nonumber \\
&&=\frac{1}{4 \pi^2} \int d \theta d r \Biggr[ \frac{1}{|\cos(\theta)|+|\sin(\theta)|} \nonumber \\
&& - \frac{r}{\sqrt{r^2 \cos^2(\theta)+1}+\sqrt{r^2 \sin^2(\theta)+1}}\nonumber \\
&& \times\Biggl(1+ \frac{1}{\sqrt{r^2 \cos^2(\theta)+1}\sqrt{r^2 \sin^2(\theta)+1}} \Biggr) \Biggr] \nonumber \\
&& = \lim_{\Lambda \rightarrow \infty} \int d \theta F_{\Lambda} (\theta) .
\end{eqnarray}

\begin{eqnarray}
&& F_{\Lambda} (\theta) \equiv \frac{1}{4 \pi^2} \int_{0}^{\Lambda} \Biggr[ \frac{1}{|\cos(\theta)|+|\sin(\theta)|}\nonumber \\
&& - \frac{r}{\sqrt{r^2 \cos^2(\theta)+1}+\sqrt{r^2 \sin^2(\theta)+1}}\nonumber \\
&& \times \Biggl(1+ \frac{1}{\sqrt{r^2 \cos^2(\theta)+1}\sqrt{r^2 \sin^2(\theta)+1}} \Biggr) \Biggr] 
\end{eqnarray}
In Fig. \ref{F}, we illustrate the function, $F_{\Lambda} (\theta)$, with different cutoffs. 
As we can see the larger cutoffs make the smaller $C_{\Delta}(0)$, even though it is positive.


\begin{thebibliography}{99}

\bibitem{imai1} Fanlong Ning, K.~Ahilan, T.~Imai1, A.~S.~Sefat, R.~Jin, M.~A.~McGuire, B.~C.~Sales, 
and D.~Mandrus, J. Phys. Soc. Jpn {\bf 78}, 013711 (2009).

\bibitem{joerg0} N.~Ni, M.~E.~Tillman, J.-Q.~Yan, A.~Kracher, S.~T.~Hannahs, 
S.~L.~Bud'ko, and P.~C.~Canfield, Phys. Rev. B {\bf 78}, 214515 (2008).

\bibitem{imai2} K.~Ahilan, F.~L.~Ning, T.~Imai, A.~S.~Sefat, M.~A.~McGuire, B.~C.~Sales, and D.~Mandrus, 
Phys. Rev. B {\bf 79}, 214520, (2009).

\bibitem{imai3} F.~L.~Ning, K.~Ahilan, T.~Imai, A.~S.~Sefat, M.~A.~McGuire, B.~C.~Sales, D.~Mandrus, P.~Cheng, 
B.~Shen, and H.-H~Wen, Phys. Rev. Lett. {\bf 104}, 037001 (2010).

\bibitem{joerg1} S. Nandi, M. G. Kim, A. Kreyssig, R. M. Fernandes, D. K. Pratt, A. Thaler, N.~Ni, S.~L.~Bud'ko, P. C. Canfield, J. Schmalian, R. J. McQueeney, and A. I. Goldman, Phys. Rev. Lett. {\bf 104}, 057006 (2010).

\bibitem{joerg2} R. M. Fernandes, D. K. Pratt, W. Tian, J. Zarestky, A. Kreyssig, S. Nandi, M. G. Kim, A.~Thaler, N. Ni, S.~L.~Bud'ko, P. C. Canfield, R. J. McQueeney, J. Schmalian, and A. I. Goldman, Phys. Rev. B {\bf 81}, 140501(R) (2010).

\bibitem{nmrjapan} Y.~Nakai, T.~Iye, S.~Kitagawa, K.~Ishida, H.~Ikeda, S.~Kasahara, H.~Shishido, T.~Shibauchi, 
Y.~Matsuda, and T.~Terashima, arXiv:1005.2853 (2010).

\bibitem{knebel} G. Knebel, D. Aoki, and J. Flouquet, arXiv:0911.5223.

\bibitem{scalapino} D.~J.~Scalapino, E.~Loh, and J.~E.~Hirsch, Phys. Rev. B {\bf 34}, 8190 (1986).

\bibitem{ChubukovLong}
Ar.~Abanov, A.~V.~Chubukov, and J.~Schmalian, Advances in Physics {\bf 52}, 119 (2003).

\bibitem{max2} M.~A.~Metlitski and S.~Sachdev, arXiv:1005.1288.

\bibitem{sachdevzhang} S. Sachdev and S.-C. Zhang, Science {\bf 295}, 452 (2002).

\bibitem{katano} S.~Katano, M.~Sato, K.~Yamada, T.~Suzuki, and T.~Fukase, Phys. Rev.
B {\bf 62}, R14677 (2000).

\bibitem{lake} B.~Lake,
H.~M.~R\o nnow, N.~B.~Christensen, G.~Aeppli, K.~Lefmann,
D.~F.~McMorrow, P.~Vorderwisch, P.~Smeibidl, N.~Mangkorntong,
T.~Sasagawa,  M.~Nohara, H.~Takagi, and T.~E.~Mason,
Nature {\bf 415}, 299 (2002).

\bibitem{jtran2} J.~M.~Tranquada, C.~H.~Lee, K.~Yamada, Y.~S.~Lee, L.~P.~Regnault, and
H.~M.~R\o nnow, Phys. Rev. B 69, 174507 (2004).

\bibitem{boris} B.~Khaykovich,
S.~Wakimoto, R.~J.~Birgeneau, M.~A.~Kastner, Y.~S.~Lee,
P.~Smeibidl, P.~Vorderwisch, and K.~Yamada,
Phys. Rev. B {\bf 71}, 220508 (2005).

\bibitem{chang1} J.~Chang,
Ch.~Niedermayer, R.~Gilardi, N.~B.~Christensen,
H.~M.~R\o nnow, D.~F.~McMorrow, M.~Ay, J.~Stahn, O.~Sobolev, A.~Hiess,
S.~Pailhes, C.~Baines, N.~Momono, M.~Oda, M.~Ido, and J.~Mesot,
Phys. Rev. B {\bf 78}, 104525 (2008).

\bibitem{chang2} J.~Chang, N.~B.~Christensen,
Ch.~Niedermayer, K.~Lefmann, H.~M.~R\o nnow, D.~F.~McMorrow, A.~Schneidewind,
P.~Link, A.~Hiess, M.~Boehm, R.~Mottl, S.~Pailhes, N.~Momono, M.~Oda, M.~Ido,
and J.~Mesot, Phys. Rev. Lett. {\bf 102}, 177006 (2009).

\bibitem{mesot3} D.~Haug, V.~Hinkov, A.~Suchaneck, D.~S.~Inosov, N.~B.~Christensen, A.~Ivanov, T.~Keller, C.~T.~Lin,
 and B.~Keimer, arXiv:1008.4298).

\bibitem{keimer} D.~Haug, V.~Hinkov, Y.~Sidis, P.~Bourges, N.~B.~Christensen, D.~S.~Inosov,
Ch. Niedermayer, P.~Bourges, Y.~Sidis, J.~T.~Park, A.~Ivanov, C.~T.~Lin, J.~Mesot, and B.~Keimer,

\bibitem{eg} E.~G. ~Moon and S.~Sachdev, Phys. Rev. B
  {\bf 80}, 035117 (2009). 

\bibitem{qcnp} S.~Sachdev, Physica Status Solidi B {\bf 247}, 537 (2010) [arXiv:0907.0008].

\bibitem{zhang}  E.~Demler, S.~Sachdev and Y.~Zhang, 
Phys. Rev. Lett. {\bf 87}, 067202 (2001); Y.~Zhang, E.~Demler, and S.~Sachdev, Phys. Rev. B. {\bf 66}, 094501 (2002). 

\bibitem{rkk1} R. K. Kaul, A. Kolezhuk, M. Levin, S. Sachdev, and T. Senthil, Phys. Rev. B  {\bf 75} , 235122 (2007).

\bibitem{rkk2} R. K. Kaul, Y. B. Kim, S. Sachdev, and T. Senthil, Nature Physics {\bf 4}, 28 (2008).

\bibitem{victor} V.~Galitski and S.~Sachdev, Phys. Rev. B {\bf 79}, 134512 (2009).

\bibitem{subir} S.~Sachdev, M. A. Metlitski, Y. Qi, and C. Xu, Phys. Rev. B {\bf 80}, 155129 (2009).

\bibitem{abanov} Ar. Abanov, A.V. Chubukov, and J. Schmalian, Adv. Phys. {\bf 52}, 119 (2003).


\bibitem{gapan1} R. Khasanov, D. V. Evtushinsky, A. Amato, H. H. Klauss, H. Luetkens, Ch. Niedermayer, 
B. B\"uchner, G. L. Sun, C.~T.~Lin, J. T. Park, D. S. Inosov, and V. Hinkov, Phys. Rev. Lett  {\bf 102}, 187005 (2009). 

\bibitem{gapan2} P. Szab\'o, Z. Pribulov\'a, G. Prist\'a\u{s}, S. L. Bud'ko, P.~C.~Canfield, and P. Samuely, 
Phys. Rev. B  {\bf 79}, 012503 (2009). 

\bibitem{mazinpm} I.~I.~Mazin, D.~J.~Singh, M.~D.~Johannes, and M.-H.~Du, Phys. Rev. Lett. 
{\bf 101}, 057003 (2008).

\bibitem{jiangping} Kangjun Seo, B.~A.~Bernevig, and Jiangping Hu, Phys. Rev. Lett. {\bf 101}, 206404 (2008).

\bibitem{kohsaka}
Y. Kohsaka, C. Taylor, K. Fujita, A. Schmidt, C. Lupien, T. Hanaguri, M. Azuma, M. Takano,
H. Eisaki, H. Takagi, S. Uchida, and J. C. Davis,
Science {\bf 315}, 1380 (2007).

\bibitem{hudson} W.~D.~Wise, M.~C.~Boyer, Kamalesh Chatterjee, Takeshi Kondo, T.~Takeuchi, H.~Ikuta, 
Yayu Wang, and  E.~W.~Hudson, Nature Physics {\bf 4}, 696 (2008).


\bibitem{vojta1} M. Vojta, Y. Zhang, and S. Sachdev, Phys. Rev. Lett. {\bf 85}, 4940 (2000); Erratum {\bf 100}, 089904(E) (2008).
\bibitem{vojta2} M. Vojta, Y. Zhang, and S. Sachdev, Int. J. Mod. Phys. B {\bf 14}, 3719 (2000).
\bibitem{vicari1} M. D. Prato, A. Pelissetto, and E. Vicari, Phys. Rev. B. {\bf 74}, 144507 (2006).
\bibitem{vicari2} A. Pelissetto, S. Sachdev, and E. Vicari, Phys. Rev. Lett. {\bf 101}, 027005 (2008).
\bibitem{kim}  E. A.~Kim, M.~J.~Lawler, P.~Oreto, S.~Sachdev, E.~Fradkin, and S.~A.~Kivelson, Phys. Rev. B {\bf 77}, 184514 (2008).
\bibitem{huh} Y.~Huh and S.~Sachdev, Phys. Rev. B {\bf 78}, 064512 (2008).

\bibitem{hinkov08a}
V. Hinkov, D. Haug, B. Fauqu\'e, P. Bourges, Y. Sidis, A. Ivanov, C. Bernhard, C. T. Lin, and B. Keimer,
Science {\bf 319}, 597 (2008).

\bibitem{taill10b}
R. Daou, J. Chang, D. LeBoeuf, O. Cyr-Choiniere, F. Laliberte, N. Doiron-Leyraud, B.~J.~Ramshaw, R. Liang,
D. A. Bonn, W. N. Hardy, and L. Taillefer,
Nature {\bf 463}, 519 (2010).

\bibitem{lawler10} M.~J.~Lawler, K.~Fujita, Jhinhwan Lee, A.~R.~Schmidt, Y.~Kohsaka, Chung Koo
Kim, H.~Eisaki, S.~Uchida, J.~C.~Davis, J.~P.~Sethna, and Eun-Ah~Kim,  Nature {\bf 466}, 347 (2010).

\bibitem{kivelson} C.~Fang, H.~Yao, W.~Tsai, J.~Hu, S.~A.~Kivelson, Phys. Rev. B {\bf 77}, 224509 (2008).

\bibitem{cenke} C. Xu, M. M\"uller, and S. Sachdev, Phys. Rev. B. {\bf 78}, 020501(R) (2008).

\bibitem{mazin} S.~A.~J.~Kimber, D.~N.~Argyriou, and I.~I.~Mazin, arXiv:1005.1761 (2010).

\bibitem{chuang} T.-M.~Chuang, M.~P.~Allan, Jinho Lee, Yang Xie, Ni Ni, S. L. Bud'ko, 
G. S. Boebinger, P.~C.~Canfield and J.~C.~Davis, Science, {\bf 327}, 181 (2010).


\end{thebibliography}
\end{document}